\documentclass[prl,reprint,superscriptaddress,amsmath,amssymb,amsfonts,showpacs]{revtex4-1}

\usepackage{graphicx}

\newcommand*{\Par}[1][C]{\ensuremath{\mathcal{#1}}}

\newcommand*{\kto}{\ensuremath{ k_{t}}}
\newcommand*{\kt}{\ensuremath{ \langle \kto  \rangle}}
\newcommand*{\kin}{\ensuremath{ k_{in}}}
\newcommand*{\ki}{\ensuremath{ \langle \kin  \rangle}}

\newcommand*{\kon}{\ensuremath{k_{out}}}
\newcommand*{\ko}{\ensuremath{ \langle \kon  \rangle}}

\newcommand*{\card}[1]{\ensuremath{f(#1)}}

\newcommand*{\pro}[1]{\ensuremath{ \mathcal{#1}}}
\newcommand*{\Puno}{\ensuremath{\pro{P}}}
\newcommand*{\PUno}[1]{\ensuremath{\Puno^{#1}}}
\newcommand*{\PUNo}[3]{\ensuremath{\PUno{#3} \! \left ( #1, #2 \right )}}

\newcommand*{\muc}{\ensuremath{ \mu_{c} }}
\newcommand*{\mucER}{\ensuremath{ \muc^{ER} }}
\newcommand*{\mucMF}{\ensuremath{ \hat{\mu}_c}}

\newcommand*{\mucS}{\ensuremath{ \mu_c^{\Par[R]}}}
\newcommand*{\mucSt}{\ensuremath{ \tilde{\mu}_c^{\Par[R]}}}
\newcommand*{\mucf}{\ensuremath{ \mu_c^{f}}}
\newcommand*{\mucL}{\ensuremath{ \mu_c^{L}}}

\newcommand*{\MIN}{\ensuremath{\bar{I}}}
\newcommand*{\Min}[2]{\ensuremath{ \MIN\!  ( \Par[#1],\Par[#2]  )}}

\newcommand*{\sca}[2][l]{\ensuremath{\mathcal{F}^{#1} \! \left (#2 \right )}}

\begin{document}

\title{Stochastic fluctuations and the detectability limit of network communities.}

\author{Lucio Floretta}
\affiliation{Laboratoire de Biophysique Statistique, Ecole Polytechnique F\'ed\'erale de Lausanne (EPFL), CH-1015 Lausanne, Switzerland}
\author{Jonas Liechti}
\affiliation{Laboratoire de Biophysique Statistique, Ecole Polytechnique F\'ed\'erale de Lausanne (EPFL), CH-1015 Lausanne, Switzerland}
\author{Alessandro Flammini}
\affiliation{School of Informatics and Computing, Indiana University, Bloomington, Indiana 47406, USA}
\author{Paolo \surname{De Los Rios}}
\affiliation{Laboratoire de Biophysique Statistique, Ecole Polytechnique F\'ed\'erale de Lausanne (EPFL), CH-1015 Lausanne, Switzerland}

\date{\today}

\begin{abstract}
We have analyzed the detectability limits of network communities in the framework of the popular Girvan and Newman benchmark. 
By carefully taking into account the inevitable stochastic fluctuations that affect the construction of each and every instance of the benchmark, we come to the conclusions that the native, putative partition of the network is completely lost even before the in-degree/out-degree ratio becomes equal to the one of a structure-less Erd\"os-R\'enyi network. We develop a simple 
iterative scheme, analytically well described by an infinite branching-process, to provide an estimate of the true detectability limit.
Using various algorithms based on modularity optimization, we show that all of them behave (semi-quantitatively) in the same way,
with the same functional form of the detectability threshold as a function of the network parameters.
Because the same behavior has also been found by further modularity-optimization methods and for methods based on different heuristics implementations, we conclude that indeed a correct definition of the detectability limit must take into account 
the stochastic fluctuations of the network construction.
\end{abstract}


\maketitle

In the process of introducing the notion of \textit{modularity} \cite{Girvan:2002p6409}, \citeauthor{Girvan:2002p6409} (GN) also proposed a method to compare the performance of \emph{community detection algorithms} \cite{Fortunato:2010p1899,Newman:2012p7307}, which is still at the basis of modern assays \cite{Lancichinetti:2008p6919,Lancichinetti:2009p6920,Lancichinetti:2009p6333}. In the GN benchmark, different algorithms are tested on a set of \textit{planted} $l$-\textit{partition models} \cite{RSA:RSA1001}.  In a nutshell, a planted $l$-partition model is a network composed of $l$ groups of $n$ vertices that are stochastically connected with each other: an edge between any two vertices within the same group is present with probability $p_{in}$, whereas an edge between any two vertices belonging to different groups is present with probability $p_{out}$.  Accordingly, the average internal degree is $\ki  = (n-1)p_{in}$, whereas the average external degree is $\ko = (l-1) n p_{out}$. To measure the extent of the community structure present in the planted $l$-partition model, it is customary to introduce the mixing parameter $\mu$ that is defined by the relations $\ko = \mu \kt$ and $\ki= (1-\mu) \kt$, where $\kt = \ki + \ko$  is the average total degree. 
Indeed, at $\mu=0$ the network has $l$ disconnected components; as $\mu$ increases, the average internal degree decreases while the average external degree toward one specific other cluster, $\ko/(l-1)$, increases until they become equal at $\mu = (l-1)/l \equiv \mucER$, when the planted $l$-partition model is practically an Erd\"os-R\'enyi graph \cite{Erdos:1959p7189}.  The performance of different algorithms is then ranked according to the value of the mixing parameter beyond which they can no longer recover the \textit{native} $l$ clusters. Thus, the GN benchmark tries to encode the simple heuristics that communities correspond to groups of vertices that are more connected with each other than with the rest of the network, although of course a rigorous univocal definition of ``more connected'' is presently lacking.

Naively, one would expect any clustering algorithm to be successful at most up to $\mucER$. However, this is not the case because the construction of a planted $l$-partition model is a stochastic process. Indeed, it might happen even at $\mu < \mucER$ that a vertex \emph{declared} as belonging to a group has fewer connections toward its putative community than toward a different one.  According to the fundamental heuristic definition of community, outlined above, this fluctuation should be interpreted as a change of membership (a \textit{relabeling}) of the vertex. Thus, we can expect that the community structure will be badly degraded when the difference between 
$\langle k_{in} \rangle$ and  $\langle k_{out} \rangle /(l-1)$ is comparable with its own statistical fluctuation, an argument that translates into the condition $\ki - \ko /(l-1) = \alpha_{l} \sqrt{\kt}$, where $\alpha_{l}$ is  a positive real constant that may depend on $l$. 
Thus, we can argue that the native community structure disappears when
\begin{equation}
\mu > \mucER \left (1 - \frac{\alpha_{l}}{\sqrt{\kt}}\right) 
\label{eq::mu_NOI}
\end{equation} 
and any algorithm trying to recover the native partition based on criteria that adhere to the heuristics should consequently fail to do so because, in essence, there is no native structure to be recognized anymore.

In order to gauge the extent to which stochastic fluctuations skew network realizations away from the native one, we have analyzed the planted $l$-partition models of the original GN benchmark ($l=4$, $n=32$, and $\kt=16$). Nodes with more connections to an outer group than to their putative one were relabeled, and the procedure was iteratively repeated until no more nodes had to change their membership. 
For each realization, we measured the similarity between the relabeled partition \Par[R]\ and the native partition \Par[N]\ using the \textit{normalized mutual information} \Min{R}{N}\ \cite{1742-5468-2005-09-P09008}. 
 \Min{R}{N}\ is defined as 
 \begin{equation*}
\Min{R}{N} \equiv \frac{2 \,I \! \left ( \Par[R], \Par[N] \right )}{H \! \left ( \Par[R] \right) +H \! \left ( \Par[N] \right) } \;,
\end{equation*}
where 
\begin{equation*}
I \! \left ( \Par[R], \Par[N] \right ) =  \sum_{\rho \in \Par[R]} \sum_{\nu \in \Par[N]} \card{\rho,\nu} \log \left ( \frac{\card{\rho,\nu}}{\card{\rho}\card{\nu}} \right) \;,
\end{equation*}
is the mutual information of \Par[R] and \Par[N], $H \! \left ( \Par[R] \right)=- \sum_{\rho \in \Par[R]} \card{\rho} \log \card{\rho}$ is the entropy of \Par[R], and $H \! \left ( \Par[N] \right)=- \sum_{\nu \in \Par[N]} \card{\nu} \log \card{\nu}$ is the entropy of \Par[N].
$\card{\rho}$ is the fraction of nodes that belong to the community $\rho$ within \Par[R], $\card{\nu}$ is the fraction of nodes that belong to the community $\nu$ within \Par[N], and $\card{\rho,\nu}$ is the fraction of nodes that are simultaneously assigned to the community $\rho$ within \Par[R] and to the community $\nu$ within \Par[N].  

\begin{figure}
\centering
\includegraphics{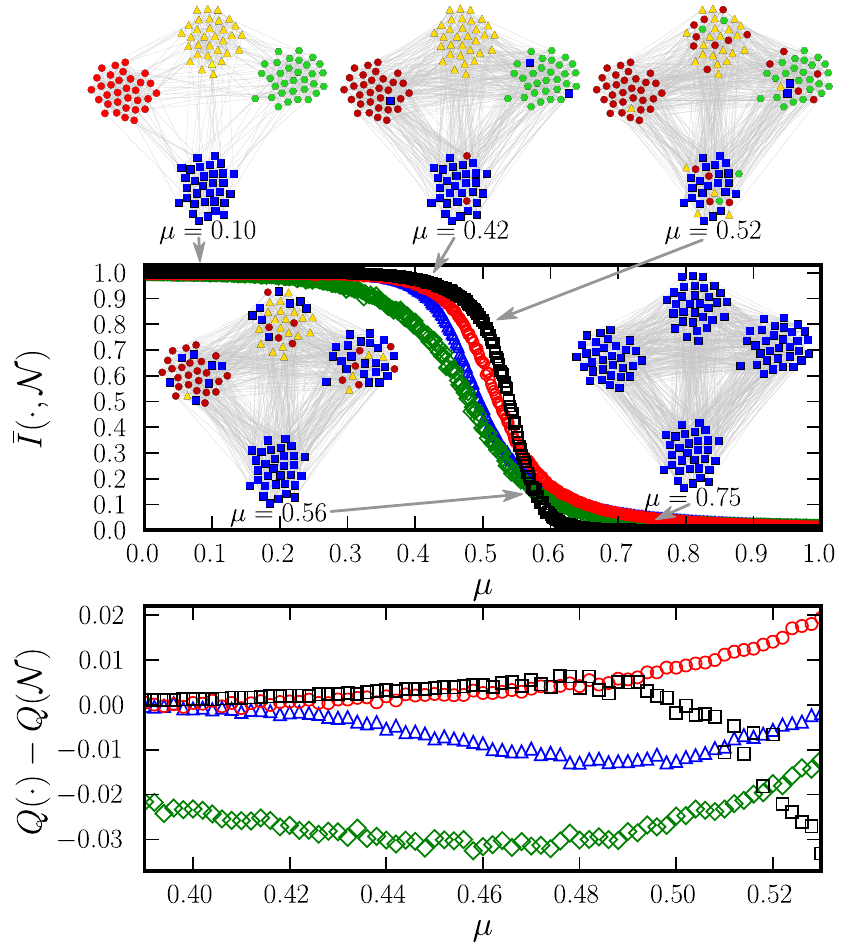}
\caption{Similarity (upper panel) along with the modularity difference (lower panel) between the relabeled partition $(\Box)$, the partitions retrieved by simulated annealing ($\circ$ in red), by fastgreedy ($\Diamond$ in green), and by the Louvain method ($\triangle$ in blue) with the native partition for the original GN benchmark ($n=32$, $l=4$, $\kt=16$). Each point is averaged over 500 realizations, the error is smaller than the marker size. Further, five networks at different $\mu$ are depicted to visualize how relabeled vertices slowly invade other putative clusters.  \label{fig::val_GN}}
\end{figure}

The outcome is shown in Fig.~\ref{fig::val_GN}: because \Min{R}{N}\  decreases as $\mu$ increases and since it goes to zero well before $\mucER = 3/4$, a labeling closer to what a zero order heuristics would suggest is already different enough from the native partition that the latter could not be detected anymore.
As modularity was conceived with the same simple heuristics in mind, we have applied three different clustering algorithms based on its maximization to the networks of Fig.~\ref{fig::val_GN}: simulated annealing \cite{Good:2010p6692}, that looks for the absolute maximum, and two \textit{greedy} algorithms with different implementations, fastgreedy \cite{Clauset:2004p6792} and Louvain method \cite{Blondel:2008p6328}. 
Interestingly, the modularity of the relabeled partition, $Q(\Par[R])$, can be larger than the modularity of the native one, $Q(\Par[N])$, in a significant region of $\mu$ values. Even more intriguingly, there the modularity found by simulated annealing roughly coincides with the modularity of the relabeled partition, $Q(\Par[R])$, and the similarity between the retrieved partition and  the relabeled partition is larger than the similarity between the retrieved partition and the putative partition, 
signaling that modularity optimization and relabeling agree on the detected deviations from the native partition. 
Greedy methods do not follow the same trend, likely because of the roughness of the modularity landscape, that hampers these algorithms for small values of the total degree \cite{Good:2010p6692}.

\begin{figure*}
\centering
\includegraphics{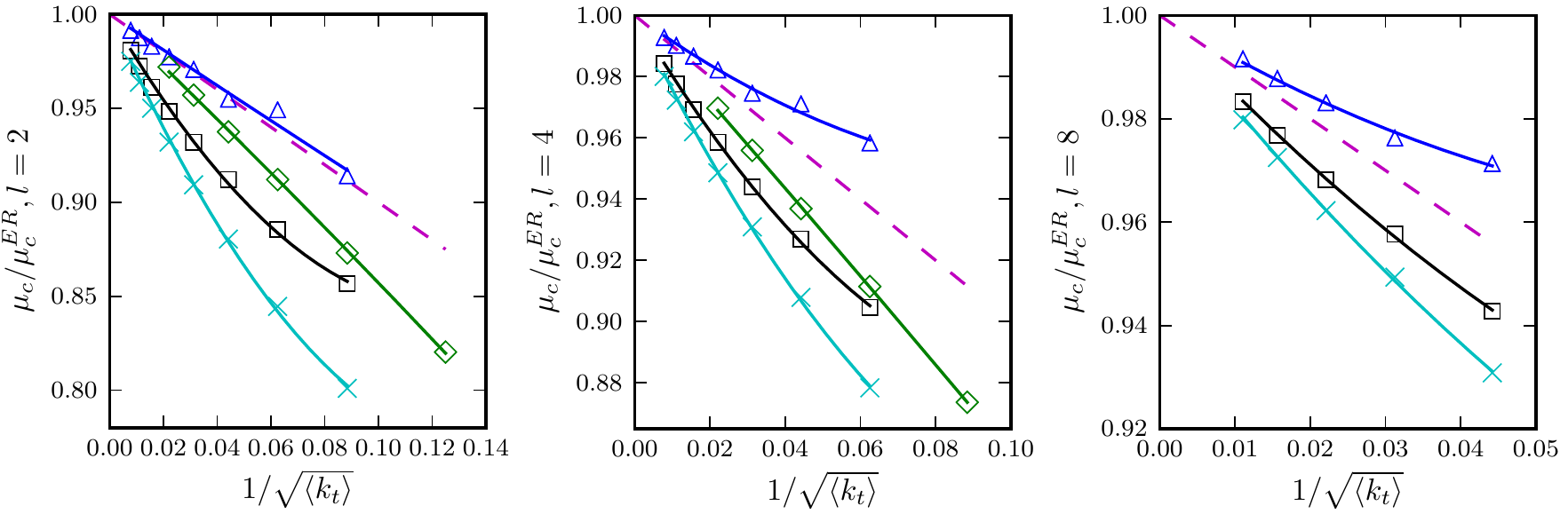}%
\caption{Comparison between the similarity thresholds \mucMF\ (Eq.~\ref{eq::mu_LN}, dashed line), \mucSt\ (Eq.~\ref{eq::condizione}, $\times$), \mucS\ ($\Min{A}{R} \approx 0.01$ , $\Box$), \mucf\ ($\Min{A}{N} \approx 0.01$ for fastgreedy, $\Diamond$), and \mucL\ ($\Min{A}{N} \approx 0.01$ for the Louvain method, $\bigtriangleup$) for $l=2$ (left panel), 4 (central panel), and 8 (right panel). In the last case, the results for fastgreedy are not available because the network sizes required to exit the glassy phase are out of its reach. The solid lines are fits of the form $\muc = \mucER \left (1 - \alpha / \sqrt{\kt} + \beta/ \kt \right )$. Errors are smaller than markers and to improve the visualization everything is divided by \mucER.}
\label{fig::threshold}
\end{figure*}

We then moved  to larger networks, for various values of the total average degrees and for different number of putative communities
($l=2,4,8$). Once again our goal was to use relabeling to discern when stochastic fluctuations completely distort the native partition.
We applied the iterative relabeling procedure outlined above, expectedly finding that
the similarity between the relabeled partition and the native one decreased as \mucER\ was approached. 
The similarity threshold, defined as the value \mucS\ of the mixing parameter where  \Min{R}{N}\ falls below 0.01, was very well described by Eq.~\eqref{eq::mu_NOI}, up to corrections of order $1/\kt$. Furthermore,  $\alpha_{l}$ approaches 1 from above as $l$ increases (Fig.~\ref{fig::threshold}, $\Box$), therefore recovering the detectability threshold
\begin{equation}
\mucMF \equiv \mucER \left (1 - \frac{1}{\sqrt{\kt}}\right) 
\label{eq::mu_LN}
\end{equation}
recently derived with statistical inference \cite{Decelle:2011p6682,PhysRevE.84.066106} and modularity \cite{Nadakuditi:2012p6840} based methods.
These results confirm thus that the stochastic fluctuations affecting the network construction are so strong
to make the native structure disappear before $\mucER$, and consistently with Eq.~\eqref{eq::mu_NOI}.

\begin{figure}
\centering
\includegraphics{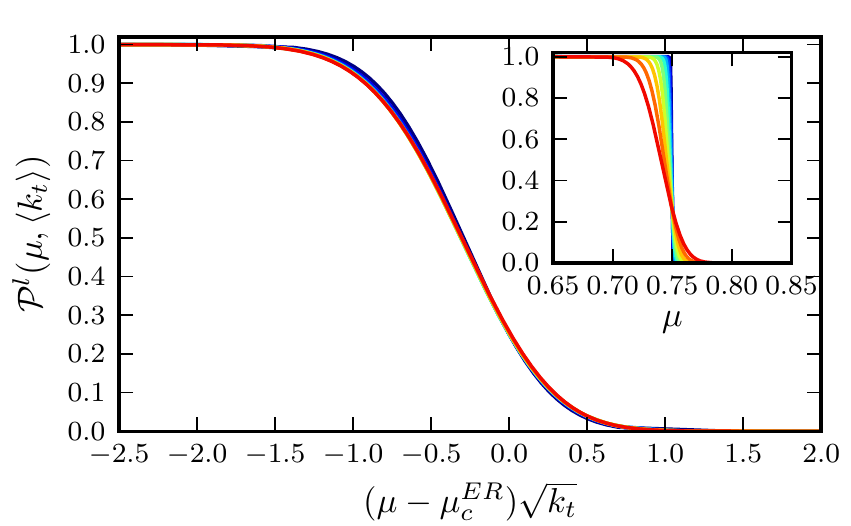}
\caption{Magnification of the data collapse of \PUNo{\mu}{\kt}{l}\ for $l=4$ and for different values of \kt\ (different colors). The inset displays the original data. \label{fig::p1_scaling}}
\end{figure}

The functional form of Eq.~\eqref{eq::mu_NOI} can be formally derived from the topology of the planted $l$-partition model and from the aforementioned definition of membership by requiring that each relabeled vertex leaves in its wake further vertices to be relabeled, thus triggering an infinite avalanche. We defined \PUNo{\mu}{\kt}{l}\ as the probability that a node has an internal degree that  is larger or equal than its number of connections with nodes of another given group. Assuming that the network is tree-like in the same mean-field spirit of \cite{Decelle:2011p6682,PhysRevE.84.066106,Nadakuditi:2012p6840},  in the limit $\kt \to \infty$ an avalanche becomes infinite at the value \mucSt\ of the mixing parameter satisfying 
\begin{equation}
\left ( \left (1 - \mucSt \right ) \kt  - 1 \right )\left ( 1 -  \PUNo{\mucSt}{\kt}{l}\right ) = 1\;,
\label{eq::condizione}
\end{equation}
as in this limit the probability that a node with a relabeled neighbor maintains its membership converges  to \PUNo{\mu}{\kt}{l} (see SI). For an infinite planted $l$-partition model 
\begin{equation}
\begin{aligned}
\PUNo{\mu}{\kt}{l}  &= e^{-\kt} \sum_{i=0}^{\infty} \frac{1}{i !}  \left ((1 - \mu) \kt \right )^{i} \\
&\mspace{105mu}\left ( \sum_{j=0}^{i} \frac{1}{j !} \left (\frac{ \mu \kt}{(l-1)} \right )^{j} \right )^{l-1} \;;
\end{aligned}
\label{eq::p1}
\end{equation}
further, it is possible, although cumbersome, to prove analytically that \PUNo{\mu}{\kt}{2}\ depends only on the rescaled mixing parameter $(\mu - \mucER) \sqrt{\kt}$ up to corrections of the form $(a + b / \kt)$ when $\kt \to \infty$ (see SI). Therefore, we made the scaling ansatz $\PUNo{\mu}{\kt}{l} = \sca{(\mu - \mucER) \sqrt{\kt}}$ which is also numerically confirmed for $l>2$ (Fig.~\ref{fig::p1_scaling} for $l=4$).

Taking these results together, the (numerical) solution of Eq.~(\ref{eq::condizione}) yields \mucSt\ as a function of \kt\ and it indeed shows that Eq.~\eqref{eq::mu_NOI} is valid up to corrections of order $1/\kt$  (Fig.~\ref{fig::threshold}, $\times$). We can again confirm that it is a general feature of the relabeled community structure to be appreciably different from the ``native'' one at some $\mu < \mucER$ (and scaling as in Eq.~\eqref{eq::mu_NOI}).

\begin{figure}
\centering
\includegraphics{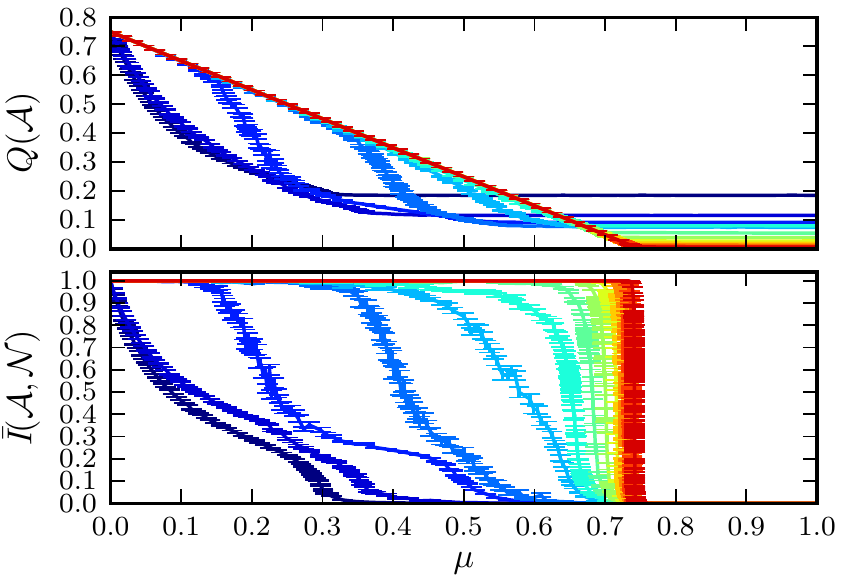}
\caption{Modularity (upper panel) of the partition \Par[A]\ retrieved by the Louvain method and its similarity index \Min{A}{N}\ with the native partition \Par[N]\ (middle panel) for the planted $4$-partition models with $n=32768$ and various $\kt$. Each data point is averaged over 100 realizations; only some errors are shown for clarity. \label{fig::blondel}}
\end{figure}

Armed with this new view of the resilience to fluctuations of the GN benchmark, we also reanalyzed the performance of modularity optimization methods fastgreedy and Louvain (we could not test simulated annealing for the sizes under scrutiny here) \cite{Lancichinetti:2009p6333}. In spite of their differences, their behavior is qualitatively similar at low \kt, when the modularity landscape
is rough, giving rise to a \textit{glassy phase} \cite{Good:2010p6692}: at low values of $\mu$ the two algorithms, that just look for local modularity maxima, are not able to find partitions with a modularity as high as the native one, which on average follows the expected mean-field value $Q^{MF} \! \left ( \mu \right ) = \mucER - \mu$ (Fig.~\eqref{fig::blondel}, upper panel; we only show the results for the Louvain method, which allows for better statistics on larger networks).
In that region, though, the native partition still fairly well corresponds to the heuristics, because there the relabeling procedure involves just a very limited number of nodes, if any (see SI). 
Also the similarity index \Min{A}{N} between the retrieved partition \Par[A] and the native partition \Par[N] drops to zero (in a rather erratic way) (Fig.~\eqref{fig::blondel}, lower panel), confirming the significant difference between the partition retrieved by the algorithms and the native one. Thus, low values of \kt\ (sparse networks) do indeed cause detectability problems much more serious than a simple shift of the threshold value of $\mu$, because of the presence of multiple competing modularity maxima.
In order to corroborate this interpretation, we initialized the Louvain method with the native planted $l$-partition, finding that the modularity of the retrieved partition never fell below its expected mean-field value (see SI).

\begin{figure}
\centering
\includegraphics{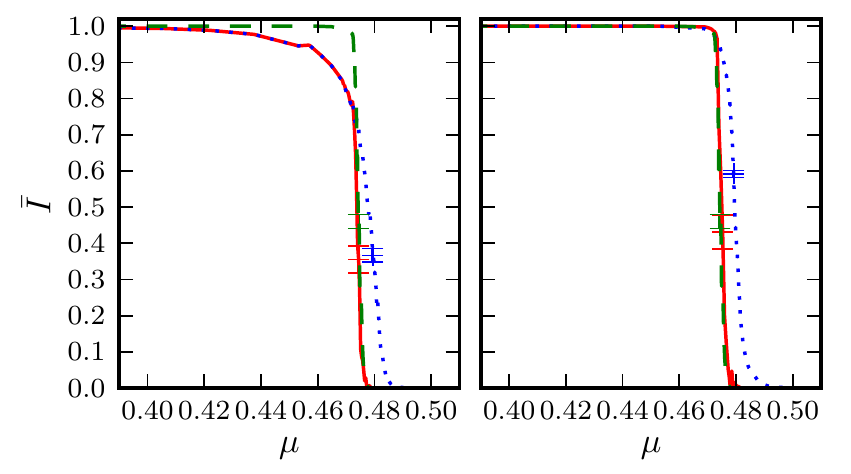}
\caption{Comparison close to the phase transition point among \Min{A}{R} (solid red line), \Min{A}{N} (dotted blue line), and \Min{R}{N} (dashed green line) for fastgreedy (left panel) and for the Louvain method (right panel) on a $2$-planted partition with $n=4096$, $\kt=2048$. For clarity, only the largest standard deviation is shown. \label{fig::comparison}}
\end{figure}

The behavior of the two algorithms differs instead outside of the glassy phase (Fig.~\ref{fig::comparison}). 
The community structure detected by fastgreedy departs from the native partition earlier than the relabeled one, implying that the algorithms does not recover the native partition even when no nodes need to be relabeled. The Louvain method finds clusters that are instead almost identical to the ones found upon relabeling, indicating that the performance of the Louvain method is optimal. 

Nevertheless, in the narrow interval defined by the dramatic drop of all similarities, the one between the relabeled partitions and the native partition, \Min{R}{N}\ decreases more rapidly and reaches zero earlier than the ones of both greedy algorithms (Fig.~\ref{fig::comparison}), and we quantified this difference by investigating in more detail  the similarity thresholds \mucf\ and \mucL\ (now defined as the value of the mixing parameter at which \Min{A}{N} drops below 0.01). We found that both \mucf\ and \mucL\ are also well fitted by Eq.~\eqref{eq::mu_NOI} with corrections of the order $1/\kt$ (Fig.~\ref{fig::threshold}, $\Diamond$ and $\bigtriangleup$, respectively). Moreover $\mucS < \mucf < \mucMF \le \mucL$ and they approach each other as $l$ increases . The differences are likely due to the different ability of each procedure to take into account higher order correlations, beyond the simple one taken into account by the relabeling procedure.

Taking together these observations, we conclude that deciding whether community detection algorithms really fail is trickier than previously believed. Indeed, in order to quantify the degree of success of a clustering method, some sort of ground truth should be known about the network under scrutiny. Since this is in general not available, various benchmarks have been developed, the simplest one being the GN, with the naive expectation that, at least for such artificial cases, the \textit{exact} solution should be known. Unfortunately, their construction is itself affected by random fluctuations that blur their structure, making it different from the putative one.
Carefully considering such statistical variations, we have shown that it is possible to change the definition of \textit{detectable region}, recognizing that fluctuations disrupt the putative partition for values of $\mu < \mucER$ and well described by Eq.~\eqref{eq::mu_NOI},
so that one should expect all community detection algorithms to be affected, qualitatively, in the same way, with differences likely to be 
imputed to different ways to implement mathematically the fundamental heuristics behind the GN benchmark.
Indeed, the finding that algorithms based on completely different principles (modularity-based \textit{vs}. statistical inference) show the same threshold beyond which a network structure becomes undetectable, must be reinterpreted
in the light of our results: beyond such threshold the real community structure simply bears no resemblance anymore with the putative one.

The present results also call for a complete reconsideration of the detectability thresholds recently derived for benchmarks richer in structure  \cite{Decelle:2011p6682,PhysRevE.84.066106,Nadakuditi:2013p7283,Peixoto:2013p7305,Radicchi:2013p7306} and of the performance of community detection algorithms tested on them \cite{Lancichinetti:2008p6919,Lancichinetti:2009p6920,Lancichinetti:2009p6333}.
On a more philosophical note, we might even relax the stringency of benchmarking protocols, and resign to the fact that clustering is,
at the end of the day, an \textit{ill posed problem} \cite{Domany:1999p7300}.

This work was financially supported by the Swiss National Science Foundation (Grant No. 200021\_125277/1).


\bibliography{lucio_clean}

\end{document}